\documentclass{vienna-conf2019}
\usepackage{graphicx}
\usepackage{hyperref}
\usepackage[]{natbib}  
\usepackage{epstopdf}

\def\BibTeX{{\rm B\kern-.05em{\sc i\kern-.025em b}\kern-.08em
    T\kern-.1667em\lower.7ex\hbox{E}\kern-.125emX}}
\bibpunct{(}{)}{;}{a}{}{,}  

\begin{document}

\TitreGlobal{Stars and their variability observed from space}


\title{The distance of the long period Cepheid RS Puppis from its remarkable light echoes}

\runningtitle{The light echoes of RS Puppis}

\author{P. Kervella}\address{LESIA, Observatoire de Paris, Universit\'e PSL, CNRS, Sorbonne Universit\'e, Univ. Paris Diderot, Sorbonne Paris Cit\'e, 5 place Jules Janssen, 92195 Meudon, France}




\setcounter{page}{237}


\maketitle


\begin{abstract}
The Milky Way Cepheid RS Puppis is a particularly important calibrator for the Leavitt law (the Period-Luminosity relation). It is a rare, long period pulsator ($P=41.5$\,days), and a good analog of the Cepheids observed in distant galaxies. It is the only known Cepheid to be embedded in a large ($\approx 0.5$\,pc) dusty nebula, that scatters the light from the pulsating star. Due to the light travel time delay introduced by the scattering on the dust, the brightness and color variations of the Cepheid imprint spectacular light echoes on the nebula. I here present a brief overview of the studies of this phenomenon, in particular through polarimetric imaging obtained with the HST/ACS camera. These observations enabled us to determine the geometry of the nebula and the distance of RS\,Pup. This distance determination is important in the context of the calibration of the Baade-Wesselink technique and of the Leavitt law.
\end{abstract}

\begin{keywords}
stars: variables: Cepheids, techniques: polarimetric, scattering, techniques:photometric, stars: distances, distance scale
\end{keywords}


\section{Introduction}

Their high intrinsic brightness and the tight relation between their period and luminosity established by the Leavitt law \citep{1908AnHar..60...87L,1912HarCi.173....1L} make long period Cepheids the most precise standard candles for the determination of extragalactic distances. Their relatively high mass ($m \approx 10-15\,M_\odot$), together with the brevity of their passage in the instability strip combine to make them very rare stars. As a result, only a handful of long-period Cepheids ($P > 30$\,d) are located within a few kiloparsecs of the Sun. RS\,Puppis (HD 68860), whose period is $P=41.5$\,d, is among the most luminous Cepheids in the Milky Way. I present in Sect.~\ref{lightechoes} the remarkable light echoes that occur in its circumstellar nebula, and how the distance to the Cepheid can be derived from their observation. Sect.~\ref{modeling} is dedicated to the modeling of the pulsation of RS\,Pup, and the calibration of the Baade-Wesselink projection factor.
 
\section{Light echoes in the nebula of RS Puppis\label{lightechoes}}
The circumstellar dusty nebula of RS Puppis was discovered by \citet{1961PASP...73...72W}. This author proposed that the presence of the dusty environment results in the creation of ``light echoes'' of the photometric variations of the Cepheid (see, e.g., \cite{2003AJ....126.1939S} for a review on this phenomenon). The variable illumination wavefronts emitted by the Cepheid propagate in the nebula, where they are scattered by the dust grains. The additional optical path of the scattered light compared to the light coming directly from the Cepheid causes the appearance of a time delay (and therefore a phase difference) between the directly observed Cepheid cycle and the photometric variation of the dust in the nebula.

The first detection of the light echoes of RS\,Pup was achieved by \citet{1972A&A....16..252H}. More than 30 years later, CCD observations were collected by \citet{2008A&A...480..167K} using the EMMI imager installed at the 3.6-m ESO NTT. These authors derived a distance based on the phase lag of selected dust knots with respect to the Cepheid. \citet{2009A&A...495..371B} however objected that the hypothesis of coplanarity of the selected dust knots was likely incorrect (see also \cite{2008MNRAS.387L..33F}), and that the determination of the 3D structure of the dust is necessary to derive the distance using the echoes.
To measure this distribution, polarimetric imaging has been shown to be a powerful technique, as applied for instance on V838\,Mon by \citet{2008AJ....135..605S} (see also \cite{1994ApJ...433...19S}).
Using this technique, the structure of the nebula was determined by \citet{2012A&A...541A..18K}, from ground-based observations with the VLT/FORS instrument. The dust is distributed over a relatively thin, ovoidal shell, probably swept away by the strong radiation pressure from the Cepheid light. These authors conclude that the dust mass in the nebula is too large to be explained by mass loss from the Cepheid itself, and that the nebula is a pre-existing interstellar dust cloud in which the Cepheid is temporarily embedded.

\begin{figure}[ht!]
 \centering
\includegraphics[width=0.9\textwidth,clip]{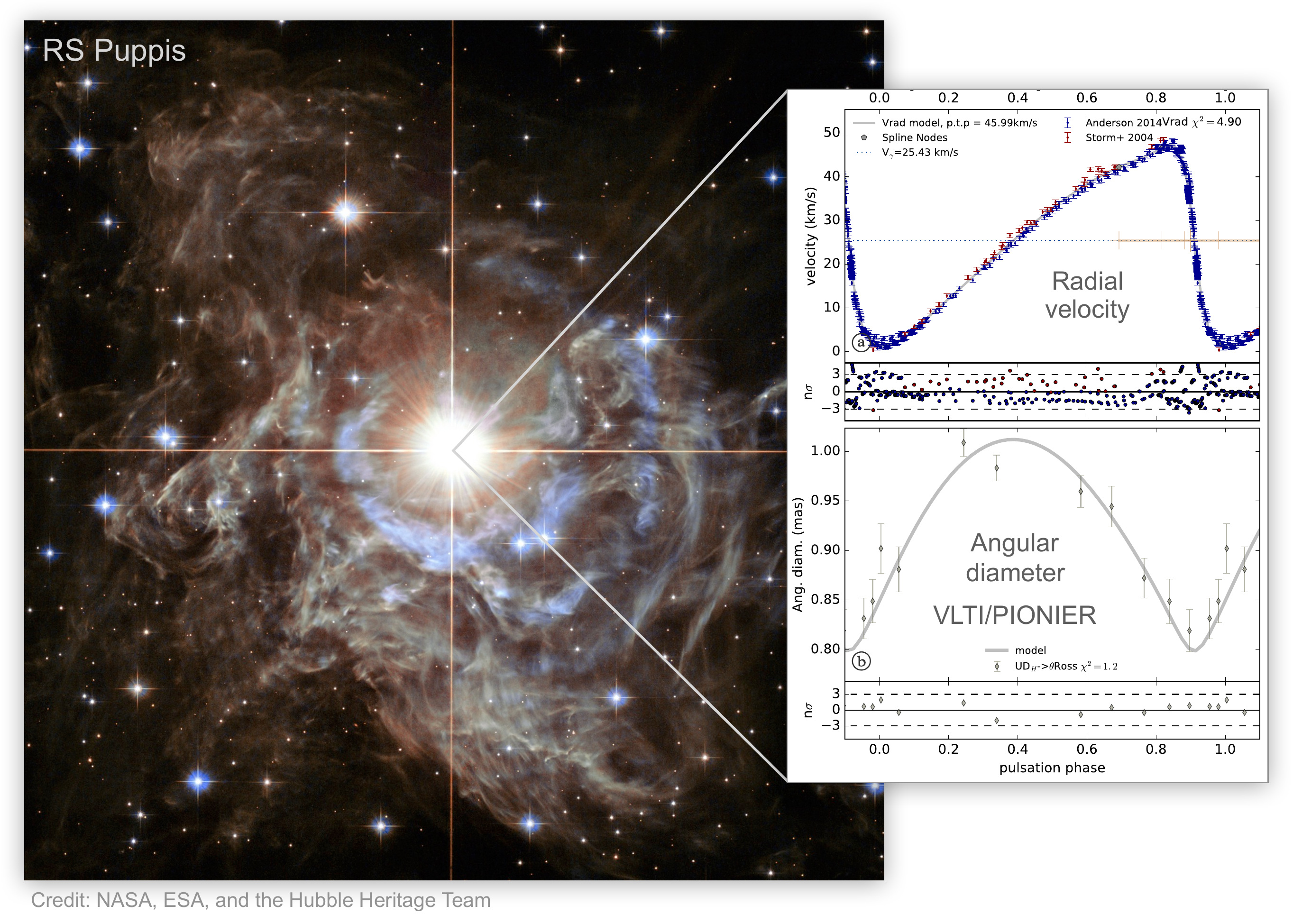}      
  \caption{{\bf Left:} HST/ACS color image of the nebula of RS Pup \citep{Kervella:2014lr}. The light echoes of the maximum light phase are visible as irregular blue rings, due to the hotter temperature of the star at its maximum light. {\bf Right:} Radial velocity and angular diameter data used for the pulsation modeling of the Cepheid \citep{2017A&A...600A.127K}.}
  \label{rspup:fig1}
\end{figure}

\begin{figure}[ht!]
 \centering
\includegraphics[width=0.4\textwidth,clip]{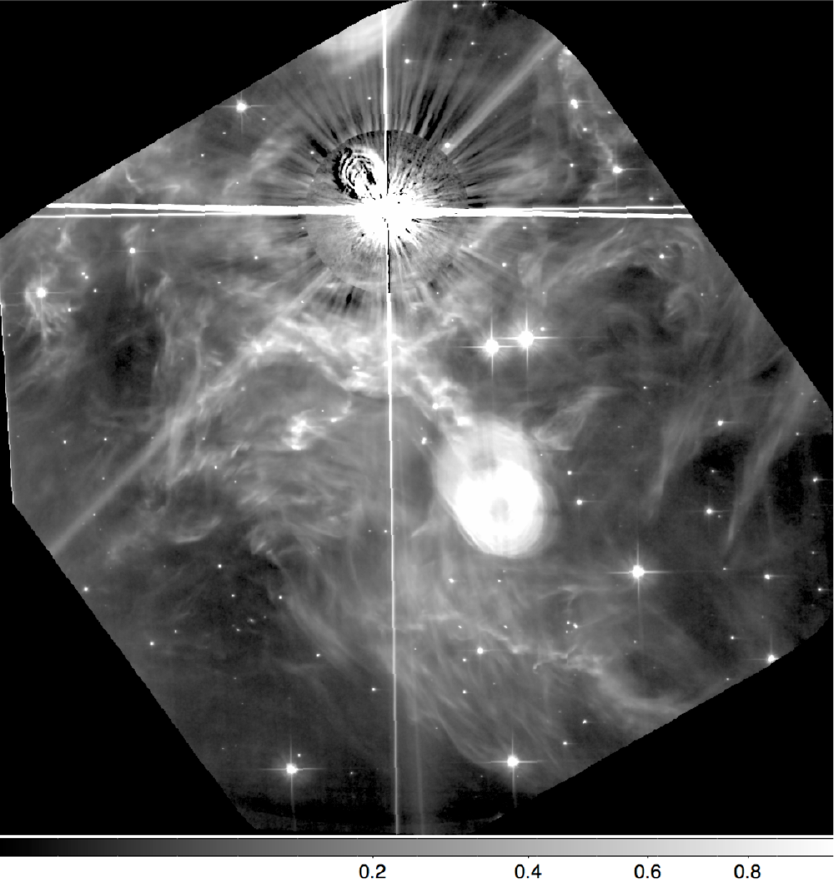}      
\includegraphics[width=0.4\textwidth,clip]{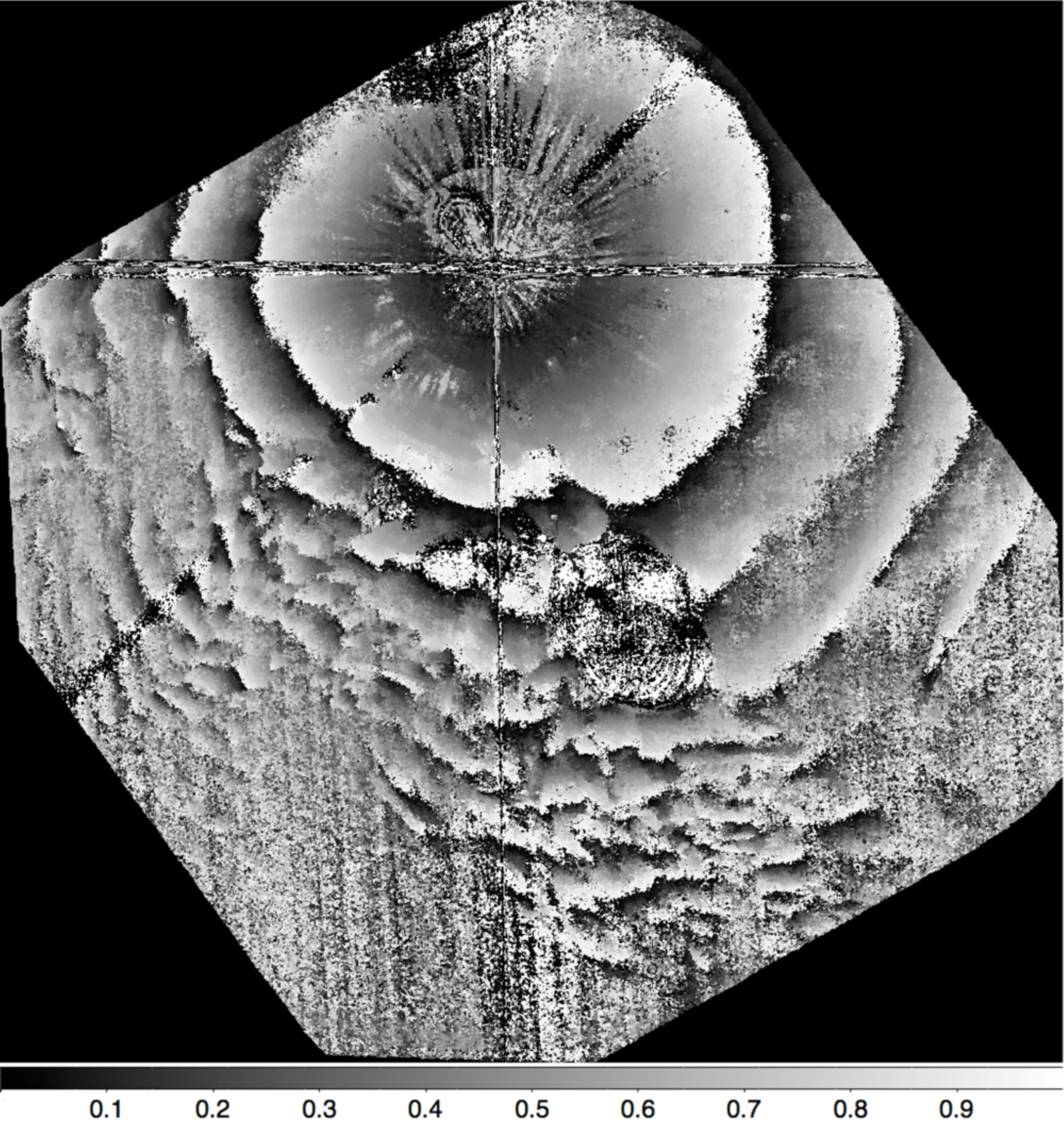}      
  \caption{{\bf Left:} Mean scattered light intensity around RS\,Pup \citep{Kervella:2014lr} (arbitrary grey scale). {\bf Right:} Phase delay of the light variation over the nebula with respect to the Cepheid cycle.}
  \label{rspup:fig2}
\end{figure}

However, the seeing-limited VLT/FORS polarimetric images could not provide a sufficiently detailed map of the structure of the nebula. A time sequence of polarimetric images of RS\,Pup was therefore obtained by \citet{Kervella:2014lr} using the HST/ACS instrument. They showed in spectacular details the structure of the nebula (Fig.~\ref{rspup:fig1}, left panel), as well as the propagation of the echoes (Fig.~\ref{rspup:fig2}; see also \url{https://vimeo.com/108581936} for a video of the light echoes), as well as the degree of linear polarization of the scattered light over the nebula.

The combination of the polarization information and the phase delay of the light echoes over the nebula result in a distance of $1910 \pm 80$\,pc ($4.2\%$), equivalent to a parallax of $\varpi = 0.524 \pm 0.022$\,mas \citep{Kervella:2014lr}. This value is significantly different from the parallax of RS\,Pup $\varpi_\mathrm{GDR2} = 0.613 \pm 0.026$\,mas (adopting a Gaia parallax shift of +29\,$\mu$as) reported in the second Gaia data release \citep{2018A&A...616A...1G}. The parallax of a field star interacting with the nebula of RS\,Pup (labeled S1, $\varpi[\mathrm{S1}] = 0.532 \pm 0.048$\,mas; \citet{2019A&A...623A.117K}) is however consistent with the light echo distance to the Cepheid.

\section{Pulsation modeling and the projection factor\label{modeling}}

The classical Baade-Wesselink (BW) distance determination technique (also known as the parallax-of-pulsation) is based on the comparison of the linear amplitude of the radius variation of a pulsating star with its angular amplitude (see, e.g., \cite{2008A&A...488...25G}). The SPIPS approach is a recent BW implementation presented by \citet{2015A&A...584A..80M}, that uses a combination of spectroscopic radial velocity, multi-band photometry and interferometric angular diameters to build a consistent model of the pulsating photosphere of the star.

A strong limitation of the BW technique comes from the fact that the distance and the spectroscopic projection factor ($p$-factor) are fully degenerate. The $p$-factor is defined as the ratio between the photospheric velocity and the disk integrated radial velocity measured by spectroscopy \citep{2014IAUS..301..145N}.
The $p$-factor is presently the major limiting factor in terms of accuracy for the application of the BW technique, and the research is particularly active on this topic \citep{2007A&A...474..975G,2011A&A...534A..94S, 2012A&A...543A..55N,2013MNRAS.436..953P,2015A&A...576A..64B,2016A&A...587A.117B,2017A&A...597A..73N,2018IAUS..330..335N}.
Using the VLTI/PIONIER optical interferometer, \citet{2017A&A...600A.127K} measured the changing angular diameter  of RS\,Pup over its pulsation cycle (Fig.~\ref{rspup:fig1}, right panel). These measurements, together with the light echo distance from the HST/ACS observations, enabled these authors to resolve the degeneracy between the distance and the $p$-factor, and obtain a $p$-factor value of $p=1.25 \pm 0.06$ for RS\,Pup.

\section{Conclusion}

The determination of an accurate distance to RS\,Pup using its light echoes is an important step toward a reliable calibration of the Leavitt law, as it is still today, even after the second Gaia data release \citep{2018A&A...616A...1G}, the most accurate distance estimate to a long period Cepheid.
Moreover, this original technique is independent from the trigonometric parallax, enabling an independent validation.
The light echo distance can also be employed to calibrate the Baade-Wesselink method, and in particular the $p$-factor. A reliable calibration of this classical technique provides a solid basis to estimate the distance of Cepheids that are too far for direct trigonometric parallax measurements with Gaia.
While BW analyses of individual Cepheids is accessible at present only up to the LMC and SMC \citep{2004A&A...415..531S,2011A&A...534A..95S,2017A&A...608A..18G}, it will soon become possible with extremely large telescopes as the E-ELT or the TMT to determine the distances of individual Cepheids in significantly more distant galaxies of the Local Group.

\begin{acknowledgements}
The research leading to these results has received funding from the European Research Council (ERC) under the European Union's Horizon 2020 research and innovation program (grant agreement No. 695099, project CepBin).
Support for Program number GO-13454 was provided by NASA through a grant from the Space Telescope Science Institute, which is operated by the Association of Universities for Research in Astronomy, Incorporated, under NASA contract NAS5-26555.
Some of the data presented in this paper were obtained from the Multimission Archive at the Space Telescope Science Institute (MAST). STScI is operated by the Association of Universities for Research in Astronomy, Inc., under NASA contract NAS5-26555. Support for MAST for non-HST data is provided by the NASA Office of Space Science via grant NAG5-7584 and by other grants and contracts.
\end{acknowledgements}

\bibliographystyle{aa}  
\bibliography{Kervella_8o02} 

\end{document}